# On Rayleigh-Taylor interfacial mixing

Evgeny E. Meshkov (1); Snezhana I. Abarzhi * (2)

Sarov Institute for Physics and Technology, National Nuclear Research University, Russia (1);

The University of Western Australia, Australia (2)

* corresponding author; email snezhana.abarzhi@gmail.com

The Rayleigh-Taylor instability develops when fluids are accelerated counter to their density gradients; intense interfacial fluid mixing ensues with time. The Rayleigh-Taylor mixing controls a broad range of processes in fluids, plasmas, materials, at astrophysical and at atomic scales. In this perspective paper we briefly review theoretical and experimental approaches, and apply theory and experiment to investigate order and disorder in Rayleigh-Taylor flows. The theory finds that properties of heterogeneous anisotropic accelerated Rayleigh-Taylor mixing depart from those of homogeneous isotropic inertial turbulence, including strong correlations, weak fluctuations, and sensitivity to deterministic conditions. The experiment unambiguously observes the heterogeneity and anisotropy of Rayleigh-Taylor mixing at very high Reynolds numbers, and the stabilizing effects of acceleration and accelerated shear on the interfacial dynamics. The theory and the experiment agree with one another and evince that Rayleigh-Taylor mixing may exhibit order and laminarize, similarly to other accelerated flows, thus opening new perspectives for research of complex processes in nature and technology.

1.  **Introduction**

We observe the Rayleigh-Taylor instability (RTI) when watching water flowing from an overturned cup [1]. RTI develops when fluids of different densities are accelerated counter to their density gradients; equivalently, when the fluid interface moves with an acceleration directed toward the denser fluid and normal to the interface [2]. The RTI leads to the growth of the interface perturbations in the acceleration direction. Intensive interfacial Rayleigh-Taylor (RT) mixing of the fluids ensues with time [3-5]. RT mixing is heterogeneous, anisotropic, and statistically unsteady process with non-local interactions among the many scales [3-5]. Its properties may depart from those of homogeneous, isotropic, local, and statistically steady turbulence [3,6]. To expand our knowledge of realistic turbulent flows beyond canonical considerations, to elaborate new methods of studies of non-equilibrium dynamics, and to better understand a broad range of RT-relevant phenomena in nature and technology, one has to grasp the fundamentals of RT mixing [3-12].

RTI and RT mixing control a broad range of processes in fluids, plasmas, materials [7]. These processes can be natural and artificial, their characteristic scales can be astrophysical and atomic, and energy densities can be low and high [3,7]. Examples include inertial confinement and magnetic fusion, formation of accretion disk and explosion of supernova, material transformation under impact and optical communications, dynamics of reactive fluids and fossil fuel recovery [7].

RT flows arising in the vastly different physical regimes exhibit a number of similar features of their evolution [1-5]. RTI starts to develop when the fluid interface and/or the flow fields are slightly perturbed near the (unstable) equilibrium [1,2]. The flow develops from an initial stage where the perturbation amplitude grows quickly (exponentially in time) to a nonlinear stage where the growth slows (to a power-law in time) [2,8]. The interface is transformed to a composition of large-scale coherent structure of bubbles and spikes (with the light (heavy) fluid penetrating the heavy (light) fluid in bubbles (spikes)), and small-scale shear-driven vortical irregular structures [2,4]. The final stage of RTI is the interfacial mixing, whose dynamics is believed to be self-similar [1-5]: In RT mixing with constant acceleration the length scale in the acceleration direction grows quadratic in time [9-12]. RT flows are usually three-dimensional (3D) [3-5,8].

RT flows exhibit a 'mixture' of multifaceted properties complementing one another [3-5,7-12]. RT mixing has certain degree of order; and yet, it is a noisy process with strong coupling of all the scales [3,8,9]. RT mixing is believed to be self-similar; and yet, it retains some memory of initial conditions [10-12]. Accelerated RT mixing is characterized by increasing values of the Reynolds number, velocity scale and energy dissipation rate – that are necessary conditions for turbulence to occur, - and yet, RT flow fields are heterogeneous, even in statistical sense, its dynamics in the acceleration direction differs from that in the normal plane, and its quantities have the mean values varying in time [3-5,7-12]. In this work we apply theory and experiment to better understand RT mixing.

RTI and RT mixing are an extreme challenge to study in their direct manifestations [7]. With nearly two hundred RT-related papers published each year in peer-reviewed journals, our knowledge of this process is still limited [3-5,7-14]. In experiments, the transient character and sensitivity of large scale dynamics to small scales and initial conditions impose tight requirements on RT flow implementation, diagnostics and control [4,5,13-17]. In simulations, the needs to accurately track interfaces and capture dissipative processes demand the use of highly accurate numerical methods and massive computations [9-12]. In theory, the necessity to identify the universal properties of asymptotic solutions, grasp symmetry of RT flows and account for their noisiness inspires the development of new methods of study of non-equilibrium, multi-scale, nonlinear, and non-local dynamics [3,5,8]. In acquiring knowledge from data, a systematic interpretation of RT dynamics from data is a challenge, and requires identification of robust parameters that should be precisely diagnosed [3-5,8-14]. For reliable quantification of power-laws describing the statistically unsteady RT mixing, the flow diagnostics demands high accuracy, high precision, high data acquisition rate, and substantial span of resolved scales in space and time [3-5,7-19].

Despite these challenges, tremendous success has been recently achieved in our understanding of RT dynamics [7-14]. In fluid experiments, the experimental design and advanced diagnostics have enabled implementations and measurements of RT mixing, including the growth-rate and the fluctuations' statistics [4,14-16]. In high energy density plasmas [5,17], certain progress has been achieved in mimicking RT-relevant astrophysical phenomena in laboratory. In simulations, the Eulerian and Lagrangian methods have provided detailed information on the dynamics and morphology of RT unstable interface and RT mixing [9-12]. Efficient computational methods have been developed to model the effect of turbulence (assuming it develops) on RT mixing, and to estimate in practice the influence of this hydrodynamic noise on parameters of RT-relevant phenomena [18,19]. In theory, significant success has been reached in the understanding of fundamental properties of RTI and RT mixing [3,5,8]. For outline of the state-of-the-art experiments, theory and simulations and for rigorous physics-based discussion of properties of RTI and RT mixing, the reader is referred to research books and review papers [3-5,7-14]. Extensive reference list can be found in recent survey [19].

Here we apply the theory and the experiment to study properties of RT dynamics, with focus on acceleration effect on order and disorder in RT mixing [3-5,8,14]. The analysis is based on group theory, identifies symmetries and invariants of RT dynamics, and finds that accelerated RT mixing has strong correlations, weak fluctuations, and sensitivity to deterministic noise; it may thus keep order [3,5,8,20-25]. The experiment applies the jelly technology to unambiguously observe the heterogeneity and anisotropy of RT mixing at high Reynolds numbers, and the rising bubbles technology to directly observe the stabilizing effects of the acceleration and accelerated shear on the interfacial dynamics [3,5,14,26,27]. The theory and the experiment agree with one another and suggest that RT mixing may keep order and laminarize, similarly to other classical accelerated flows, such as flows in curved pipes and accelerated boundary layers [28,29]. The other data available indicate that for precise quantification of RT mixing a substantial span of resolved scales is required, to be achieved in the future [16,30].

## 2. Theoretical and experimental approaches

### 2.1 Theory

#### 2.1.1 Governing equations

RT dynamics of ideal fluids is governed by the conservation of mass, momentum and energy:

$$\partial\rho/\partial t + \partial\rho v_i/\partial x_i = 0, \quad \partial\rho v_i/\partial t + \sum_{j=1}^{3}\partial\rho v_i v_j/\partial x_j + \partial P/\partial x_i = 0, \quad \partial E/\partial t + \partial(E+P)v_i/\partial x_i = 0 \quad (1a)$$

Here $x_i$ are the spatial coordinates with $\{x_1, x_2, x_3\} = \{x, y, z\}$, $t$ is time, $\{\rho, \mathbf{v}, P, E\}$ are the fields of density $\rho$, velocity $\mathbf{v}$, pressure $P$, energy $E = \rho(e + \mathbf{v}^2/2)$, and $e$ is specific internal energy [31]. In the bulk of the heavy (light) fluid the flow fields are $\{\rho, \mathbf{v}, P, E\}_{h(l)}$. At the interface, for immiscible fluids the fluxes of mass, normal and tangential components of momentum and energy obey the conditions

$$[\mathbf{v}\cdot\mathbf{n}] = 0, \quad [P] = 0, \quad [\mathbf{v}\cdot\boldsymbol{\tau}] = any, \quad [W] = any \quad (1b)$$

Here $[...]$ denotes the jump of functions at the interface; $\mathbf{n}$ and $\boldsymbol{\tau}$ are the normal and tangential unit vectors of the interface; $W = e + P/\rho$ is specific enthalpy. A spatially extended flow is periodic in the $(x, y)$ plane normal to the $z$ direction of gravity $\mathbf{g}$, $|\mathbf{g}| = g$, $\mathbf{g} = (0, 0, -g)$, and has no sources [SI]:

$$\mathbf{v}|_{z\to+\infty} = 0, \quad \mathbf{v}|_{z\to-\infty} = 0 \quad (1c)$$

Initial conditions at the interface close the set of the governing equations [3,8,31]. In the presence of kinematic viscosity $\nu$ and other non-ideal effects Eqs.(1) are modified [31]. The rigorous theories and models [3,5,8,32-37] have solved the problem Eqs.(1) in well-defined approximations and have described the data with nearly same sets of parameters.

#### 2.1.2 Theoretical approaches

In the linear regime of RTI, small perturbations of the flow field grow with time, $\sim exp(t/\tau)$, where $\tau \sim \sqrt{\lambda/Ag}$ is time-scale, $\lambda$ is the initial perturbation wavelength, $A = (\rho_h - \rho_l)/(\rho_h + \rho_l)$ is the Atwood number [1,2,31]. One can account for effects of viscosity, surface tension, compressibility, thermal conductivity, magnetic fields on RTI growth-rate [5,8,32-34], These effects usually stabilize small scales with $\lambda < \lambda_c$, where $\lambda_c$ is a critical wavelength, and induce a characteristic length-scale $\lambda_m$, at which the maximum growth-rate is achieved, $\lambda_m > \lambda_c$, e.g., $\lambda_m \sim (\nu^2/g)^{1/3}$ [32]. For small (large) wavelengths, $\lambda/\lambda_m \to 0(\infty)$, the RTI growth-rate is small compared to that at $\lambda_m$ [32]. To derive RTI growth-rate for initial perturbations with various symmetries, group theory is employed [8,22].

As time progresses, the exponential growth is superseded by a time-dependent power-law. The weakly-nonlinear RTI in ideal fluids has been studied theoretically and empirically with focus on the effect of initial spectra on the growth-rate [3,5,35]. The models have found that standard perturbation approaches work well only at early times and for small amplitudes [35]. Furthermore, the nonlinear RT dynamics has a non-local character and is accompanied by singularities [3,5,8]. The non-local and singular character of the RTI is exhibited in, e.g., multiplicity of nonlinear regular asymptotic solutions [3,8]. To solve the boundary value problem Eqs.(1) and to describe the nonlinear dynamics of large-scale coherent structures in RTI, group theory is applied [3,5,8,20-22]. This approach accurately identifies the local properties of RT-unstable interface, such as regular asymptotic solutions describing RT bubbles, and the global properties, such as classification of interactions and formation of patterns [3,8,20-22].

The interactions of RT bubbles may lead to the growth of the structure wavelength(s) and the bubble merge, thus triggering transition to self-similar RT mixing [10-12,22]. This transition has been investigated by the seminal model [10,12,36] that has excellently agreed with data. The pioneering model [11,37] has accurately interpolated the RT mixing dynamics. By assuming that RT mixing is similar to canonical turbulence, several models have applied the RT mixing growth-rate in turbulent scaling laws to enable computations of turbulence effect on RT mixing [18,19].

Group theory of RT mixing [3,5,8,23-25] has been fully consistent with the merger and interpolation models [10-12,36,37], has been harmonious with canonical turbulence theory [6,31], and has been linked to turbulence models [18,19]. By scrupulously analyzing symmetries and invariants of RT dynamics, group theory has found the new properties of RT mixing [3,24].

2.2 Experiment

2.2.1 Challenges and capabilities

It is easy to overturn a cup of water to observe the RTI [1,2]. It is an extreme challenge to reliably implement RT flows in a laboratory, and to systematically probe and gather precise and accurate data on RT dynamics in experiment [4,13,14]. On the side of implementation, RT flows are sensitive to initial and boundary conditions at the interface and the outside boundaries, to acceleration history, fluids' properties and experimental bias at large and at small scales [4,13-16]. On the side of diagnostics, many powerful methods have been developed for probing fluid flows [13]. They range from measurements of average velocities and pressure differences to the Schlieren and shadowgraph techniques, thermal anemometry and laser Doppler velocimetry, planar laser induced fluorescence and particle image velocimetry [13]. These methods have substantially advanced our knowledge of canonical turbulence [38]. Their capabilities in the RT flows are somewhat bounded by the resolution limitations and the flow sensitivity to the measurement technology [4,13-16].

On the data gathering side, RT flows are characterized by statistically unsteady transports of mass, momentum, and energy, and thus demand highly resolved spatial and temporal measurements of the flow fields [3-5,13-16]. This, in turn, requires substantial advancements of data gathering techniques that are applied in canonical turbulence, where it is often sufficient to take one (or, a few) point(s) measurements of a temporal dependence of a flow field and transform it to a spatial scaling by using Taylor's hypothesis [13,38]. On data interpretation side, similarly to canonical turbulence, RT dynamics is characterized by power-laws, and their accurate quantification requires substantial span of resolved scales [13-16,38].

Tremendous success has been recently achieved in implementation and diagnostics of RTI and RT mixing in the state-of-the-art experiments [4,14,15-17]. Yet the problem is challenging, and question 'What are the qualitative and quantitative properties of Rayleigh-Taylor flows?' remains well open [14].

### 2.2.2 Experimental approaches

In this work we present the series of qualitative experiments in a broad range of setups [4,5,14, 26,27,39-44] [SI]. Some of them have been conducted independently from the theory; some are inspired by the theory. The experiments are designed to ensure repeatable observations unambiguously answering the principal questions: Is interfacial RT mixing heterogeneous and anisotropic? What is the influence of acceleration on order and disorder in RT mixing? How is RT mixing related to other accelerated flows?

Our experimental approaches include the jelly method, and the rising bubbles method. These methods are proven to be informative, robust and efficient. The flow is monitored by high-speed film and video recording and/or by a camera with an open shutter and a pulsed light source [14,26,27,39-43].

<u>Jelly experiments</u>: The jelly method has been principally developed by Meshkov, Rogachev, Zhidov, Nevmerzhitsky and colleagues in 1980s [39] and has been actively used since then [14,39-43]. It allows a broad range of setups and high repeatability for each setup. Our experiments use a jelly of aqueous gelatin solution. These jellies are made by using a standard technique. The jelly strength depends on the solution concentration. For small concentration ~4-5%, the jelly strength is $\sim 10\text{kPa}$ [SI]. The flows are accelerated by the gas pressure that is substantially greater than the jelly strength. To produce this pressure, we use either a compressed gas, or the products of detonation of acetylene-oxygen mixture. The resulting pressure is over $\sim 10^2$ greater than the jelly strength. Under this pressure, the jelly behaves as an incompressible fluid. The jelly's transparency enables the use of optical methods [14,39-43].

In jelly experiments, we use a device in the form of a miniature 'gun' with a channel of square section ($4 \times 4\text{cm}^2$). The channel is assembled from the Plexiglas plates that are $2\text{cm}$ thick. The jelly layer is cast into the hole of one of the plates and is accelerated by the gas pressure. In experiments with gas pressure produced by detonation products, initially, at temperature $300^\circ\text{K}$ and pressure $10^5\text{Pa}$, an acetylene - oxygen mixture has density $1350\text{kg}/\text{m}^3$ and detonation velocity $2450\text{m}/\text{s}$. The detonation is initiated by the electric spark(s) at one (several) point(s). After the detonation and shock waves damping, the detonation products in the chamber have pressure $\sim 1.68\text{MPa}$, temperature $\sim 3800^\circ\text{K}$, and adiabatic index $1.27$ (Gerasimov). Typical experimental values are: the speed of $4.5\text{cm}$ layer is $\sim 70\text{m}/\text{s}$, the acceleration $g$ is $(3\text{-}7) \times 10^4 \text{ m}/\text{s}^2$, the Atwood number is $A \sim 1$, the run-time $T$ of the RTI and RT mixing is $(2.5\text{-}4) \times 10^3 \mu\text{s}$. The Reynolds number Re is estimated as $1.4 \times 10^5 \text{ - } 3.2 \times 10^6$ [14,39-43] [SI]. For experiments with compressed gas, the values are smaller [SI]. In the case of compressed gas the accelerations are more uniform, when compared to the case of detonation products. The observed dynamics is essentially inviscid and incompressible [14,39-43] [SI]. The experiment applies the jelly to control the initial conditions. For random initial conditions, other liquids can be used [14]. Major qualitative result of the experiments - direct observation of heterogeneity and anisotropy of RT mixing at high Reynolds numbers - holds true with the number variations [SI].

<u>Experiments with bubbles</u>: The raising bubble method includes experiments with pop-up air bubble in water (fluids with $A \sim 1$) and water bubble in salted water (fluids with $A \ll 1$) [26,27,44]. A pop-up air bubble is formed when a shell of an inflated rubber ball is placed at the bottom of a vertically located water channel and then collapsed. The destruction of the ball shell is carried out at its pole by a needle passing through the hole through which the ball is filled with air [26]. A floating water bubble is formed in a similar manner in the channel with salt solution [27]. The results of video registration of such flow in many experiments show that, after being pierced, the rubber shell breaks into 2-3 large fragments shrinking under the action of force directed along the shell surface. The combination of a light mass of the rubber shell and a considerable force stretching it (when the bubble is being filled) causes the shell to compress rapidly, with the remnants of the shell shrinking over the surface of the bubble. Such sliding motion of the shell remnants causes small perturbations of the surface [26,27] [SI].

### 3.    Results

#### 3.1 Theory

##### 3.1.1   Linear RT dynamics

In RT dynamics, the perturbations are a superposition of standing waves [1-3]. For periodic spatially extended RT flows, theory of discrete groups can be applied [3,8]: The RT dynamics is invariant with respect to a discrete spatial group $\mathbf{G}$, whose generators are translations in the plane, rotations and reflections [SI]. The groups relevant to structurally stable RT dynamics must have anisotropy in the acceleration direction and inversion in the normal plane, such as groups of hexagon p6mm, square p4mm, rectangle p2mm in 3D, and group pm11 in 2D [3,8] [SI]. Irreducible representations of a group are applied to describe RT flow fields [3,5,8,20-22]. For small perturbations problem Eqs.(1) is linear, and its solution is straightforward [3,8] [SI].

##### 3.1.2   Nonlinear RT dynamics

Nonlinear RT flows consists of large-scale coherent structures, whose dynamics is potential, and small-scale shear-driven interfacial vortical structures [3,8]. For the large-scale dynamics, by applying the group irreducible representations, the flow fields are expanded as Fourier series; a spatial expansion is further made of Eqs.(1) in a vicinity of a regular point at the interface, e.g., the bubble tip [8,20,21] [SI]. Governing equations Eqs.(1) are then reduced to a dynamical system in terms of moments and surface variables. Group theory is further applied to solve the closure problem, find regular asymptotic solutions for the dynamical system, study their stability, and identify properties of the nonlinear RTI [3,8,20,21,45]. Some details are given in [SI].

Briefly, for the nonlinear coherent RT dynamics, the regular asymptotic solutions form a continuous family. The family solutions converge with the increase of the approximation order. The number of the family parameters is set by the flow symmetry. This multiplicity is associated with the non-local and singular character of interfacial dynamics, and is due to interfacial shear. The solutions stability is analyzed. The fastest stable solution is identified as physically significant; this solution has the invariant that is a function of wavelength $\lambda$ and amplitude $h$ and its derivatives [3,8,20,21,45] [SI].

For group p4mm, to the first order, for family solutions the bubble velocity $v$ depends on its principal curvature $\zeta$, $\zeta \in (-|\zeta_{cr}|, 0)$, with wavevector $k = 2\pi/\lambda$, as

$$v/\sqrt{g/k} = \sqrt{2A}\sqrt{(-2A\zeta/k)(9 - 64(\zeta/k)^2)}\left(\sqrt{-48(\zeta/k) + A(9 + 64(\zeta/k)^2)}\right)^{-1}$$

The shear near the bubble tip can be defined as $\Gamma = \partial(\mathbf{v}_h - \mathbf{v}_l)_{x(y)}/\partial x(y)$. For $\zeta \in (-|\zeta_{cr}|, 0)$, shear $\Gamma$ is 1-1 function on curvature $\zeta$. The fastest stable solution is the Atwood bubble with the velocity and curvature $(v_A, \zeta_A)$. It has the invariant $v_A^2/\left((g/k)(8|\zeta_A|/k)^3\right) = 1$ [3,8,20,21,45] [SI].

For these regular asymptotic solutions there is effectively no motion of the fluids away from the interface, there is intense motion of the fluids near the interface, shear is present at the interface leading to formation of the interfacial vortical structures, Figure 1. The invariant properties of the physically significant solution imply that the wavelength $\lambda$ and amplitude $h$ both contribute to the nonlinear RT dynamics. Furthermore, in RT flows, the 3D coherent structures tend to conserve an isotropy in the plane and have a discontinuous 2D-3D dimensional crossover [3,8,20,21,45] [XX].

3.1.3 Pattern formation and transition to mixing

For pattern formation in RT flows, group theory [22] is fully consistent with the seminal model [10,12,36], suggesting that under modulations a transition may occur from a periodic structure to a super-structure with larger wavelengths. To quantify the bubble merge' and the transition to self-similar RT mixing, the model [10,12,36] must be applied. This model accounts for the multi-scale character of the RT dynamics, accurately describes the dynamics of the bubble size and amplitude, and excellently agrees with experiments and simulations.

3.1.4 RT mixing

<u>Balances</u>: In RT mixing, group theory is implemented in the momentum model, whose equations have the same scaling invariance as the governing equations [3,5,23-35]. In the momentum model, the dynamics of a parcel of fluid undergoing RT mixing is governed by a balance per unit mass of the rates of momentum gain, $\tilde{\mu}$, and momentum loss $\mu$, as $\dot{h} = v$ and $\dot{v} = \tilde{\mu} - \mu$, where $h$ is the (vertical) length scale along the acceleration $\mathbf{g}$, $v$ is the corresponding velocity, $\mu(\tilde{\mu})$ is the magnitude of the rate of loss (gain) of specific momentum in the acceleration direction [3,5,23-25]. The rate of loss (gain) of specific momentum $\mu(\tilde{\mu})$ is associated with the rate of loss (gain) of specific energy $\varepsilon(\tilde{\varepsilon})$ as $\mu = \varepsilon/v$ ($\tilde{\mu} = \tilde{\varepsilon}/v$). The rate of energy gain is $\tilde{\varepsilon} = fgv$, $f = f(A)$, with $gf \to g$ re-scaled hereafter. The rate of energy dissipation is $\varepsilon = Cv^3/L$, as in canonical turbulence [6,31]. Here $C$ is the drag coefficient, and $L$ is the length scale for energy dissipation. This scale can be horizontal, $L \sim \lambda$, vertical, $L \sim h$, or a combination of scales, $L \sim L(\lambda, h)$ [3,5,8]. By applying Lie groups, the solution is sought.

For $L \sim \lambda$ the gains and losses are fully balanced, $\tilde{\mu} = \mu$ ($\tilde{\varepsilon} = \varepsilon$), and the dynamics is steady, $h \sim t\sqrt{g\lambda}$ [3,23-25]. For $L \sim h$, the rates of the gain and loss of specific momentum (energy) are imbalanced, $\widetilde{\mu} \neq \mu$ $(\widetilde{\varepsilon} \neq \varepsilon)$. The imbalance leads to self-similar RT mixing with $h \sim gt^2$ [3,23-25]. The imbalance may occur due to the growth of horizontal scale $\lambda \sim gt^2$, as the model [10,12,36] predicts. It may occur when amplitude $h$ is the dominant scale for energy dissipation. According to data, the imbalance is small, $(\widetilde{\mu} - \mu)/\widetilde{\mu} \ll 1$ [9,10-12,36].

RT mixing has new macroscopic, scaling, invariant and correlation properties [3,5,23-25].

<u>Macroscopic properties</u>: In RT mixing, gravity is the only energy source, and the position of the center of mass of the fluid system depends on time. Canonical turbulence decays unless it is driven, and requires an external source supplying energy at a constant rate [6,31]. In RT mixing, drag $C$ is rather large $C \sim 3 \div 10$ [10-12,24,25]; in canonical turbulence it is $C \sim 1$. The large drag suggests that accelerated RT mixing may laminarize, similarly to laminarizations of other accelerated flows that have been discovered for flows in curved pipes and accelerated boundary layers [28,29] [SI].

<u>Symmetries</u>: RT mixing has a number of symmetries, and is invariant with respect to scaling transformation, similarly to canonical turbulence. Yet, symmetries of RT mixing are distinct from those of turbulence [3,24]. Particularly, in canonical turbulence, the invariant of scaling transformation is the rate of dissipation of specific kinetic energy $\varepsilon \sim v^3/L$ [6,31]. Its invariance is compatible with existence of inertial interval and non-dissipative energy transport [6,23,31,38]. In RT mixing, the invariant of scaling transformation is the rate of loss of specific momentum in the acceleration direction $\mu \sim v^2/L$ [3,23,24]: The momentum and energy are gained and lost at any scale, the rates of loss (gain) of specific momentum $\mu(\widetilde{\mu})$ are invariant, and the rates of energy dissipation (gain) are time-dependent, $\varepsilon(\widetilde{\varepsilon}) \sim g^2 t$ [SI].

<u>Scaling and correlations</u>: In RT mixing the invariance of the rate of momentum loss $\mu$ leads to the scaling of velocity $v_l/v \sim (l/L)^{1/2}$ and $N$th order structure function $\sim (l\mu)^{N/2}$. In canonical turbulence the invariance of $\varepsilon$ leads to the velocity scaling $v_l/v \sim (l/L)^{1/3}$ and the $N$th order structure function $\sim (l\varepsilon)^{N/3}$. From these scaling exponents, the velocity correlations are stronger in RT mixing than in turbulence. In RT mixing, the Reynolds number $Re = vL/\nu$ is scale-dependent, $Re \sim g^2 t^3/\nu$, and the local Reynolds number $Re_l = v_l l/\nu$ scales as $Re_l/Re \sim (l/L)^{3/2}$ leading to viscous scale $l_\nu \sim (\nu^2/\mu)^{1/3}$ with $l_\nu \sim \lambda_m$ [3,23-25]. In turbulence, $Re_l/Re = (l/L)^{4/3}$ and $l_\nu \sim (\nu^3/\varepsilon)^{1/4}$ [6,31]. Span of scales increases with time in RT mixing $L/l_\nu \sim gt^2 (\mu/\nu^2)^{1/3}$, and is constant in turbulence $L/l_\nu \sim L(\varepsilon/\nu^3)^{1/4}$ [SI].

<u>Fluctuations</u>: Fluctuations are essential for turbulent systems, and their strength is expected to exceed the deterministic noise, i.e., the noise caused by the initial, external and experimental conditions. In canonical turbulence the invariance of energy dissipation rate $\varepsilon \sim v^3/L$ leads to diffusion scaling law for velocity fluctuations, $v \sim t^{1/2}$ [6,31]. In RT mixing, the invariance of the rate of momentum loss $\mu \sim v^2/L$ leads to ballistic dynamics, $v \sim t$ [3,5,23-25]. To compare the strengths of turbulent and deterministic fluctuations, consider two parcels of fluids entrained in the flow with a time-delay $\widetilde{\tau}$. In canonical turbulence, turbulent velocity fluctuations $\sim (\varepsilon v \widetilde{\tau})^{1/3}$ are substantially greater than the parcels' relative deterministic velocity $\sim (\varepsilon \widetilde{\tau})^{1/2}$; the ratio of the fluctuating and the mean velocity velocities is constant $\sim (\varepsilon \widetilde{\tau}/v^2)^{1/3}$ [6,31]. In RT mixing, turbulent velocity fluctuations $\sim \mu \widetilde{\tau}$ are comparable to the parcels' relative deterministic velocity $\sim (\widetilde{\mu}-\mu)\widetilde{\tau}$; the ratio of the fluctuating and mean velocities decays with time $\sim (\widetilde{\tau}/t)$ [3,5,24]. We see that turbulence is a stochastic process with self-generated fluctuations insensitive to deterministic noise [6,31]. In RT mixing, fluctuations are sensitive to deterministic noise, and their strength decays with time [3,5,24]. This suggests that RT mixing may laminarize [SI].

<u>Dimensional analysis based spectra</u>: In canonical turbulence, the invariance of energy dissipation rate leads to spectral density $E$ of kinetic energy fluctuations $E \sim \varepsilon^{2/3} k^{-5/3}$ [6,31]. For statistically unsteady RT mixing, spectra may be a challenge to rigorously define. The dimensional analysis-based power-law spectrum for kinetic energy fluctuations is steeper in RT mixing than in Kolmogorov turbulence, $E \sim \mu k^{-2}$ [3,24].

Note that in the balanced RT dynamics with $L \sim \lambda$ and moderate Reynolds number, a disordered state may appear for noisy initial conditions, with the span of scales $\sim (\lambda/\lambda_m)^{9/8}$ [3,5,24,25]. The accelerated ordered and chaotic disordered state may co-exist [3,5,24] [SI].

3.2 Experiment

The jelly experiments study whether RT mixing is (hetero) homogeneous and (an) isotropic [4,14,39-43].

3.2.1 Sensitivity of RT mixing to deterministic conditions

<u>Periodic perturbation</u>: Figure 2a shows the RTI development for a layer of jelly that is placed in a container and accelerated by the compressed gas [41] in a sample case of a two-dimensional (2D) flow with a single mode small amplitude sinusoidal initial perturbation. The interface perturbation first grow symmetrically. When the perturbation amplitude is ~40% of its wavelength, the perturbation becomes asymmetric, and the large-scale coherent structure of bubble and spikes appears. In the nonlinear RTI, the bubbles have constant velocity, $\sim 0.23\sqrt{g\lambda}$ and the spikes accelerate, in agreement with theory for $A \sim 1$ [3,8,14,20].

<u>Bubbles competition and merge</u>: Figure 2b illustrates the flow transition to the accelerated RT mixing via the growth of horizontal scales. The initial perturbation consists of $10 \times 10$ hemi-spherical hollows periodically placed on the jelly's layer surface. The acceleration is set by the detonation products pressure. In the linear and nonlinear regimes, the dynamics is coherent, the bubbles move with the same constant velocity, and the spikes accelerate. Bubbles' interaction is triggered by large-scale modulations (due to the acceleration non-uniformities by point detonation [SI]): Some of the bubbles merge and grow faster than their neighbors. With time, the bubbles competition increases, and the flow becomes more disordered. These experiments are in excellent agreement with the merger model [10,12,36] and with group theory [3,5,22]. The acceleration non-uniformities and fields' perturbations are discussed in [SI].

The excellent agreement of the experiments with the rigorous theories and models illustrates that the observed dynamics is essentially incompressible and inviscid [14,39-43] [SI].

<u>Localized perturbations</u>: RT dynamics of localized perturbations has somewhat different properties [14,40]. Figure 3a presents the development of RT mixing for a localized initial perturbation in the form of a hemi-spherical hollow at the unstable surface of the jelly layer. With time, the perturbation grows in the amplitude and widens laterally acquiring a nearly-spherical shape [14,41]. Figure 3b illustrates the development of RT mixing for a periodic initial perturbation in the form of the same $5 \times 5$ hollows uniformly located on the jelly layer surface. The perturbation amplitude grows, whereas the bubbles size in the transverse direction practically does not change and bubbles' merge does not occur. Note the development of small scale structures on the surface near a local disturbance in Figure 3a and the almost complete suppression of such structures by the growing periodic perturbation in Figure 3b.

### 3.2.2 Heterogeneity and anisotropy of the RT mixing

<u>Planar geometry</u>: Figure 4 shows the development of RT mixing at a flat boundary of two layers of jelly that are separated by an air gap. The lower layer is accelerated by the pressure $\sim 1.5\text{MPa}$ of the detonation products in the $4 \times 4 \text{cm}^2$ chamber which is located in the lower part of the channel. The detonation is initiated synchronously by electric sparks at $8 \times 8$ uniformly positioned points at the bottom of the chamber. The initial perturbation at the surface of the lower layer has two components: the roughness of the layer surface and the perturbations of the detonation wave. The jelly's upper layer is accelerated by the pressure of air which is compressed by the flying lower layer. The initial perturbation at this surface is set only by the surface roughness [14,39-43].

<u>Convergent geometry</u>: Figure 5 shows the development of RT mixing for a flow at a cylindrical interface in the regimes of explosion and implosion. Here, $1\text{cm}$ jelly ring is placed between two Plexiglas plates. In Figure 5a,b, the inner volume of the ring is filled with acetylene-oxygen mixture. Its detonation is initiated by an electric spark at the chamber axis. The expansion of the ring and the development of RT mixing at the inner surface of the ring are registered by a camera with an open shutter in a darkened room when the experimental device is illuminated by a flash of light. In this case, the initial perturbation is the surface roughness of the ring. In Figure 5c, the experimental assembly is consisted of two Plexiglas plates and a cylindrical shell between them; a jelly ring is placed inside the shell and concentrically to the shell. The volume between the shell and the jelly ring is filled with acetylene-oxygen mixture; its detonation is initiated by electric sparks at $40$ points that are uniformly located along the inner surface of the shell. The detonation products pressure leads to the ring collapse. RT mixing development at the outer surface of the ring is recorded by a high-speed camera. In this case, the initial perturbation is the surface roughness of the ring and the detonation wave perturbations [14].

<u>Interfacial density jump</u>: While these experimental setups vary significantly (including types of initial perturbations, accelerations, flow geometries) their results have an important qualitative feature in common, Figure 2-5. Particularly, RT mixing zone consists of two parts – the first dark, opaque, dispersed part adjacent to the gas, whose pressure accelerates the liquid (initially – jelly) layer, and the second part in the form of a transparent bright bubbles penetrating into the liquid (jelly). This is lucidly seen in the expanding ring experiment, Figure 5 [4,14]. The direct observation of the light coloration of the bubbles and their transparency implies that the bubbles are filled with the luminous detonation products having a very high temperature. Thus, at the dome of each bubble there is a density jump at the interface of the liquid (initially – jelly) and the gaseous detonation products, that is at the liquid-gas interface. Furthermore, this jump is the jump that has existed at the interface initially. The surface of the bubbles penetrating the liquid (jelly) is the strongly deformed initial surface (between the detonation products and the jelly). The density jump at the surface of the bubble domes prevents the erosion of this liquid-gas interface. It is required for development of accelerated RT mixing.

Our jelly experiments study fluids in liquid-gas systems with high Atwood numbers, $A \approx 1$. Similar results are obtained for gas-gas systems with low Atwood numbers, $A << 1$, with slightly modified mechanism of the density jump formation [4]. For a gas-gas interface, a lighter gas penetrates into the mixing zone as a mixture with a heavier gas, and the mixing zone development is carried out by the spikes of a heavy gas penetrating the mixing zone [4].

### 3.2.3  Effect of the acceleration on laminarization of RT mixing

The rising bubble experiments study stabilizing effects of acceleration and accelerated shear [26,27,44].

<u>High Atwood number</u>: The stability of the dome surface of a large air bubble emerging in the water is a long-standing problem [46]. This stability is usually associated with the effects of surface tension and bubble curvature. We show here that the bubble dome stability can be explained by the action of the accelerated shear flow of water over the surface of the air bubble [4,14,26]. In these experiments the Atwood numbers are high, $A \approx 1$. In Figure 6a, an air bubble is formed in a square section channel filled with water, after the inflated rubber shell is broken and is pierced from the inside with a needle (slide 1). The shell is tightened for $2-3\text{ms}$ and then the air bubble begins to rise from the resting state (slide 2). At the initial moment, the water in the channel is also at rest. At the same time, the RTI starts to develop at the bubble dome and the perturbations (created by the sliding ruptured rubber shell) remain at the bubble surface (slide 2). Then, as the bubble rises, after being accelerated for a short time, it moves at a constant speed. Yet, there is an accelerated water flow along the curved surface of the air bubble dome, and the growth of the perturbations at the surface of the air bubble dome is suppressed almost immediately (slides 3-5). Note that the observed stabilization mechanism is solely due to the accelerated motion of the fluid along the curved surface of the bubble dome, similarly to the flows in curved pipes [28]. In fact, this is the only difference between the bubble states in the slides 2 and 3!

<u>Low Atwood number</u>: Figure 6b shows the results of similar experiments for the low Atwood numbers $A << 1$ [27]. These are experiments with water bubble rising in salt solution, the Atwood number is $A = 0.007$ for salt concentration $15\text{g}/1$. In this experiment, the surface tension effect is negligible on the surface of the bubble dome at the water-salted water interface. Yet, at the bubble dome, an accelerated shear flow arises, and the instability does not develop. Note that at the lateral surface of the bubble, where the shear flow is nearly steady, intense development of the Kelvin-Helmholtz instability is observed [27].

### 3.3 Properties of RT dynamics

Our theory solves the boundary value and initial value problems for the linear and nonlinear RTI, and analyzes symmetries and invariants for RT mixing, Eqs.(1). For RTI, the principal results are: the multi-scale character of nonlinear coherent dynamics, to which both the wavelength and amplitude contribute; the tendency of RT flows to keep isotropy in the plane normal to the acceleration; the discontinuity of the dimensional crossover. For RT mixing, the principal results are: the invariance of the rate of momentum loss, new symmetries and scaling transformations, and order of RT mixing, due to its strong correlations, weak fluctuations, and sensitivity to deterministic conditions [3-5,8,23,24].

Our experiment directly observes RT mixing in a broad range of setups, and evinces that at high Reynolds numbers the interface between the fluids does not become smeared during the development of RT mixing [14]. The interface is strongly deformed. It is the strongly deformed initial interface between the fluids. The constant density jump is preserved at the interface at the advanced stages of RT mixing. The existence of this jump is required for the accelerated RT mixing to continuously develop. It ensures that the width of RT mixing zone increases in time at an increasing rate. The acceleration and accelerated shear, in turn, preserve the density jump at the strongly deformed interface and stabilize the dynamics [4,5,14,26,27,39-44] [SI].

The theory and the experiment excellently agree with one another lucidly show: RT mixing is heterogeneous and anisotropic. Its properties depart from the properties of homogeneous and isotropic turbulence. Accelerated RT mixing may exhibit order and laminarize, similarly to other classical accelerated flows [3,5,8,14,28,29].

In addition to qualitative results, the theory identifies quantitative benchmarks of RT dynamics. At present, they are rarely investigated. The existing state-of-the-art fluid experiments studying the effect of turbulence on RT mixing are focused on fluctuations spectra [16]. The experiments observe that RT mixing spectra are steeper than those in canonical turbulence. This indicates a clear trend toward group theory results [16]. Substantial dynamic range can help to precisely quantify RT mixing power-laws, to be achieved in the future [13,16].

### 4. Discussion

We have applied theory and experiment to study Rayleigh-Taylor flows, with focus on the effect of acceleration on order and disorder in RT mixing. Our analysis has employed group theory and has found that properties of heterogeneous, anisotropic and statistically unsteady RT mixing depart from those of canonical turbulence: The RT dynamics is multi-scale; the invariant of RT mixing is the rate of momentum loss; RT mixing flow fields exhibit strong correlations, weak fluctuations, and sensitivity to deterministic conditions. Our experiments has applied the jelly technology to unambiguously observe the heterogeneity and anisotropy of RT mixing at high Reynolds numbers, and the rising bubbles technology to directly observe the stabilizing effects of the acceleration and accelerated shear. The theory and experiment agree with one another. We find that accelerated RT mixing may keep order and laminarize.

Our theory is harmonious with models [10-12] and turbulence theory [6,31], and is linked to turbulence models [18,19,25]. Our experiment is congruous with other experiments in RT flows [15-17] and accelerated flows [28,29]. Our jelly experiments have one of largest Reynolds numbers in laboratory studies of RT flows [4,5,14-17]. Our major qualitative result - heterogeneity and anisotropy of RT mixing - holds true with the Reynolds number variations Figure 2-5 [4,14]. Details of dynamics are sensitive to deterministic conditions [4,14]. Our theory support our experiments [3,5,8]. Our results clearly indicate that new developments are in demand to study properties of RT mixing in realistic environments, and to better understand the interplay of acceleration, interfacial dynamics and scale coupling at microscopic (kinetic) and macroscopic (continuous) scale.

One possible research area with many applications in nature and technology is RT mixing with variable acceleration [5]. For instance, in supernova, these studies are required to better understand whether turbulence may appear in blast-wave-driven RT dynamics and whether mechanisms other than turbulence may exist to enable energy accumulation and nuclear synthesis at small scales [47]. In inertial confinement fusion (ICF), these investigations are in need to find new opportunities for control of RT mixing in high energy density plasmas (HEDP) [48].

Our theory suggests the following scenario for the evolution of RTI and RT mixing: Initially, small-scale structures at the interface grow quickly. The nonlinear dynamics is multi-scale, with the wavelength and amplitude contributing. In the accelerated RT mixing, momentum is imbalanced. The Reynolds number increases, yet, the accelerated RT mixing may keep order (due to strong correlations, weak fluctuations and sensitivity to deterministic noise) [3,5,8,20-25]. Our experiment supports this scenario [4,14,26,27,39-44]. If the scenario holds true for variable accelerations, then in ICF, it may be worth to scratch the target in order to pre-impose proper deterministic conditions to gain better control on the RT mixing and to achieve a more ordered plasma flow [3,5,24]. This may help the existing methods that are focused on fine polishing of the ICF target to fully eliminate the RTI (assuming that RTI, once it appears, cannot be controlled, and leads to fully turbulent RT mixing) [48].

The other possible research area is RT dynamics in miscible fluids. It is often believed that RT mixing of miscible fluids results in canonical turbulence [19,30,49]. In our experiments on RT mixing in miscible fluids, the interface is observed to be important whether the fluids are gases or liquids [4,14]. In the theory, the interface evolution with the interfacial scalar transport demands the solution of boundary value problem, and may result in new hydrodynamic instability and new mechanism of (de) stabilization of the interface with the interfacial mass flux [SI]. For simulaitons of RT mixing in miscible fluids, highly accurate numerical methods with the interface tracking are important [9-12].

Our theory and experiment indicate that depending on deterministic flow conditions a disordered (quasi-turbulent) and an ordered (laminar) dynamics may co-exist in RT flows, especially at moderate-to-high Reynolds numbers [3,5,23,24] [SI]. Studies of other flows (such as unsteady turbulent boundary layers, turbulent jets, and unstably stratified free shear flows) at various Reynolds numbers, the Atwood and Schmidt numbers, and acceleration patterns may shed further light on the interplay of acceleration and turbulence [16,29,49].

In studies of turbulence effect on RT mixing, spectra are common to diagnose [16,30]. In turbulence, these measurements are a reliable information source [38]. In RT mixing, mathematically, spectra may be a challenge to rigorously define due to statistically unsteady dynamics [31]. Physically, canonical turbulence can be viewed as a stochastic process (insensitive to deterministic conditions) transferring energy from external source to random velocity fluctuations [31]. In RT mixing acceleration is the only energy source, and fluctuations appear to be ballistic and 'deterministic' [3,24] [SI]. Future studies are required to fully understand the complexity and stochasticity of RT flows, Figure 2-6.

Our theory identifies a number of benchmarks to quantify RT dynamics [SI]. Some of them can be obtained from existing data, e.g., steeper than Kolmogorov spectra [16]. Some others require substantial improvements of diagnostics and data gathering techniques to be able to measure with high accuracy and precision, in space and in time, the multiple and time-dependent scales in a single experimental run, and to augment it with high reproducibility of the acceleration history and accurate control of the initial and boundary conditions and other macroscopic parameters [3,4,13,14]. Modern technologies can provide the state-of-the-art fluid experiments with principal improvements in precision, dynamic range, reproducibility, motion-control accuracy, data-acquisition rate and information capacity to better understand RT flows in realistic environments and to unambiguously compare the experimental results with the rigorous theory and the simulations [13,14] [SI] [SIE].

We have considered the RTI and RT mixing in their direct manifestations and have found that the accelerated heterogeneous anisotropic interfacial RT mixing may keep order. These results are consistent with classical observations of the laminarization of accelerated flows and the intermittency and multi-fractality of canonical turbulence in the presence of interfaces [28,29,38]. Further investigations are required to fully understand the interplay of turbulence, acceleration, and interfacial dynamics in broad parameter regimes. The problem of water flowing from an overturned cup is a source of inspiration for researchers in science, mathematics and engineering, and is well open for a curious mind.

**Acknowledgements**
SIA appreciates support of the US National Science Foundation and the University of Western Australia.

## 6. Figure Captions

Figure 1: One parameter family of regular asymptotic solution in the nonlinear RTI with symmetry group p4mm in incompressible immiscible ideal fluids for some Atwood numbers. Dependence of the bubble velocity on (a) the bubble curvature and (b) the interfacial shear. Qualitative velocity field in laboratory reference frame in (c) the volume and (d) the plane with the interface marked by a dashed curve [3,8].

Figure 2: Development of the initial periodic (a) 2D perturbation with wavelength 8mm accelerated at $\sim 10^4$m/s$^2$ by compressed air; (b) 3D perturbation accelerated by detonation products [14].

Figure 3: Evolution of (a) localized initial perturbation in the form of a hemispherical hollow and (b) periodic initial perturbation in the form of 25 = 5x5 hemispherical hollows on the unstable surface of the accelerated jelly layer [14].

Figure 4: The development of RT mixing at the unstable interface of two layers of jelly accelerated in a square-section channel. The lower layer is accelerated by the pressure of the detonation products (DP), and the upper layer - by the pressure of air compressed by the accelerated lower layer [14].

Figure 5: RT mixing at the inner surface of the jelly ring, under the pressure of the detonation products of the acetylene-oxygen mixture. (a) Preview frame: Jelly ring is placed between two plates of Plexiglas; the inner volume of the ring is filled with a mixture of acetylene and oxygen; the detonation is initiated at the center by a spark. (b) Snapshot of a flying ring at about a time when the RT mixing front comes out onto the ring outer surface. (c) Collapse of jelly ring under the pressure of products of detonation (D) initiated at 40 points on the inner surface of the cylindrical case (outside the frame). In the initial stage of the collapse of the shell, RT mixing (RTM) zone is formed at the outer boundary, and in the deceleration stage at the inner boundary [14].

Figure 6: (a) Snapshots of high-speed video recording of the rise of air bubble in water in a channel with 11x11cm$^2$ square cross section from the resting state [26]. (b) The rise of water bubble in salted water (A=0.007). The bubble dome surface is stable, due to the accelerated shear; the Kelvin-Helmholtz instability develops intensely on the lateral surface of the bubble [27].

Figure 1

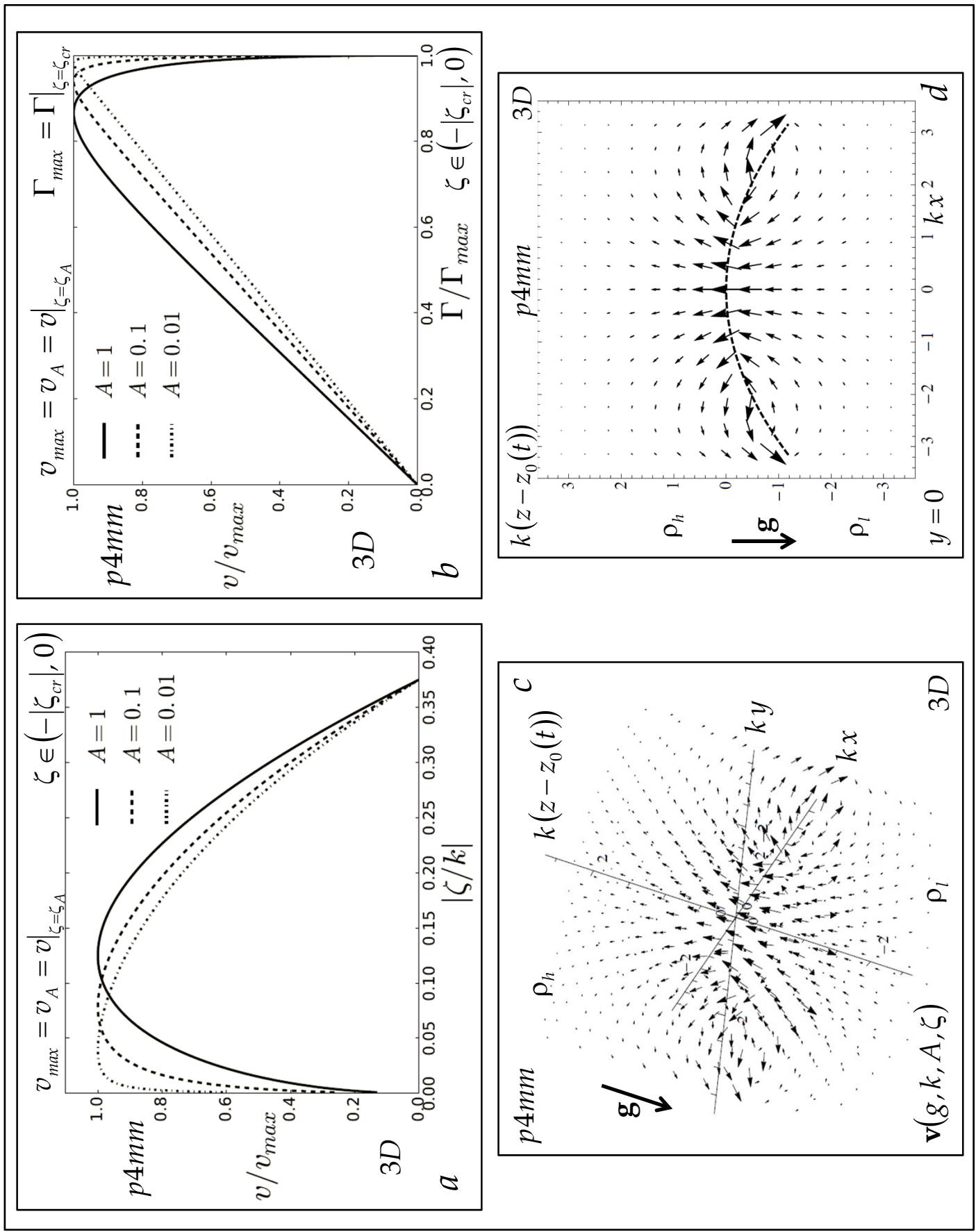

Figure 2

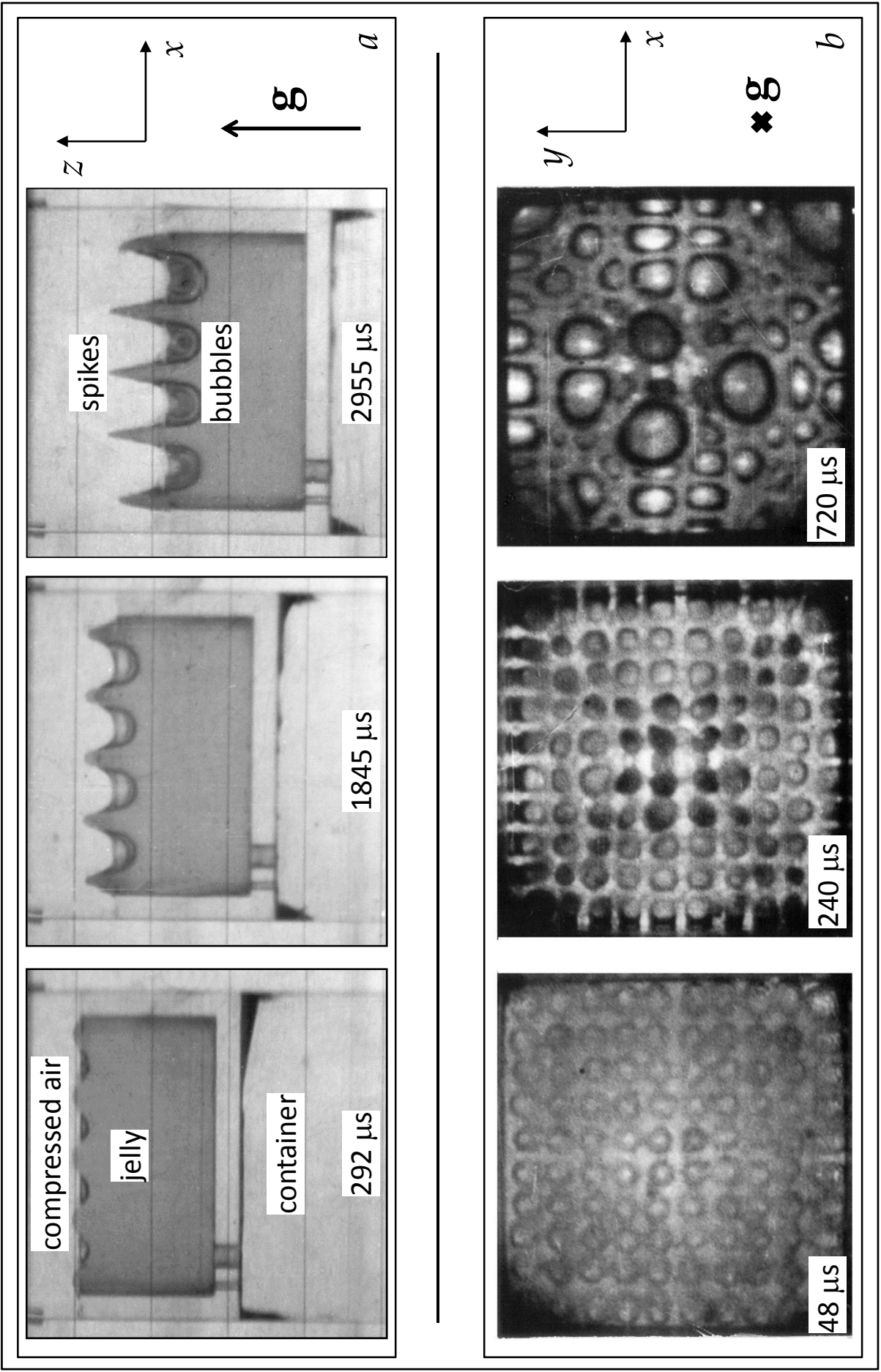

Figure 3

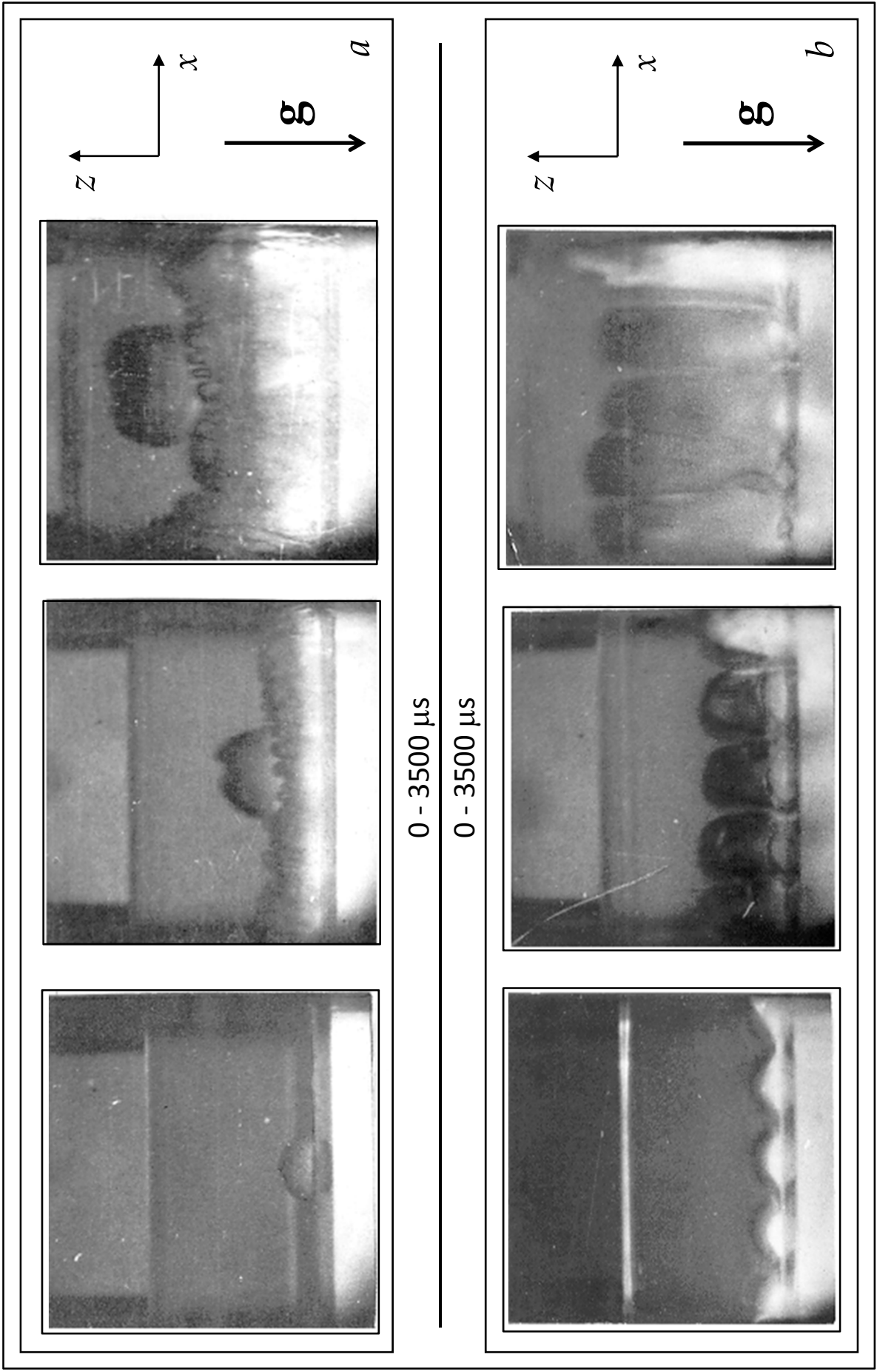

Figure 4

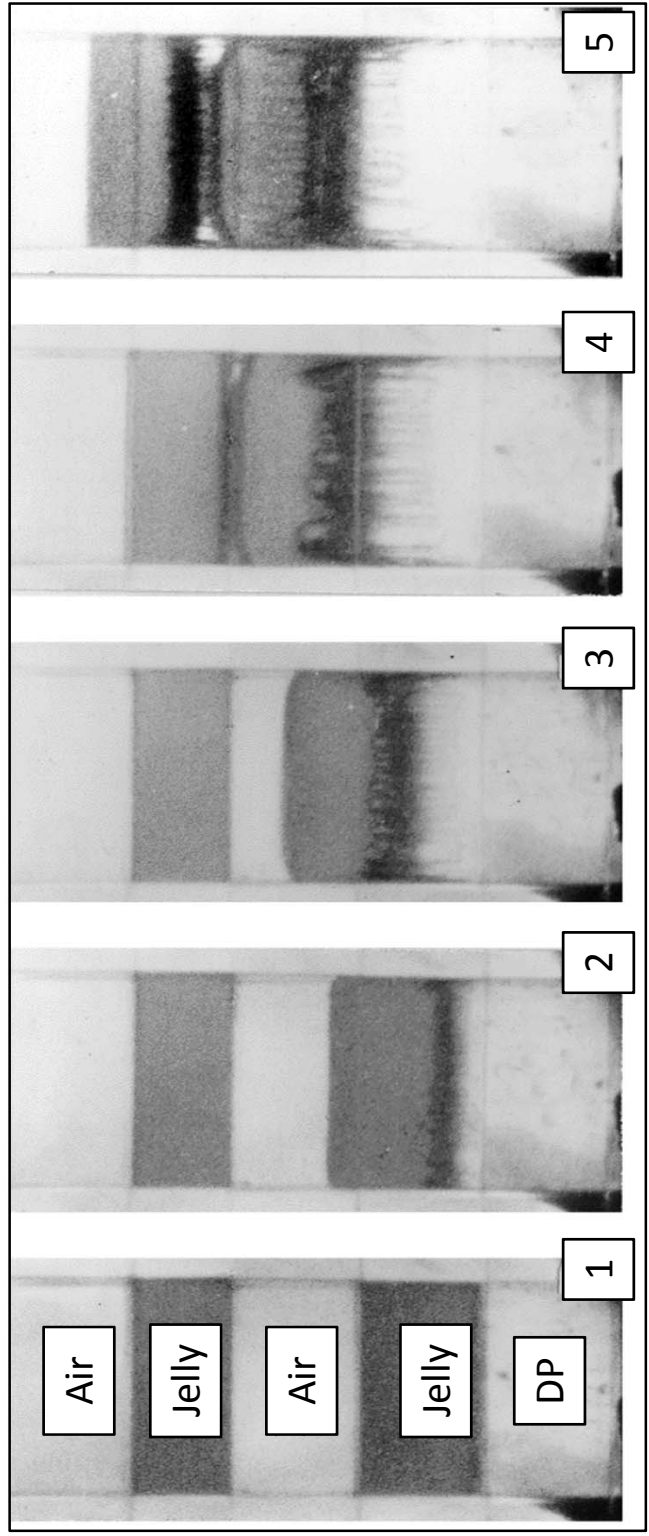

Figure 5

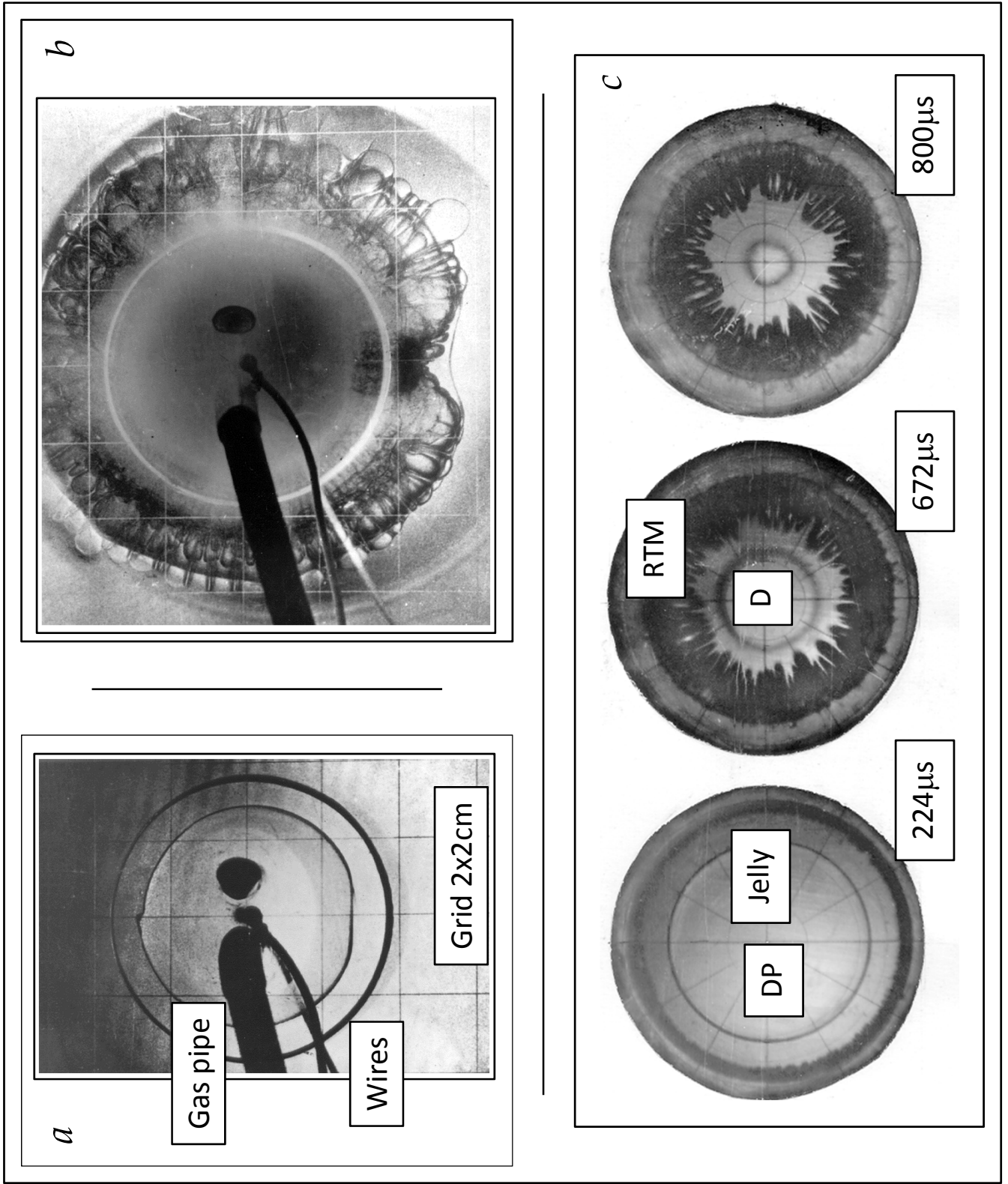

Figure 6

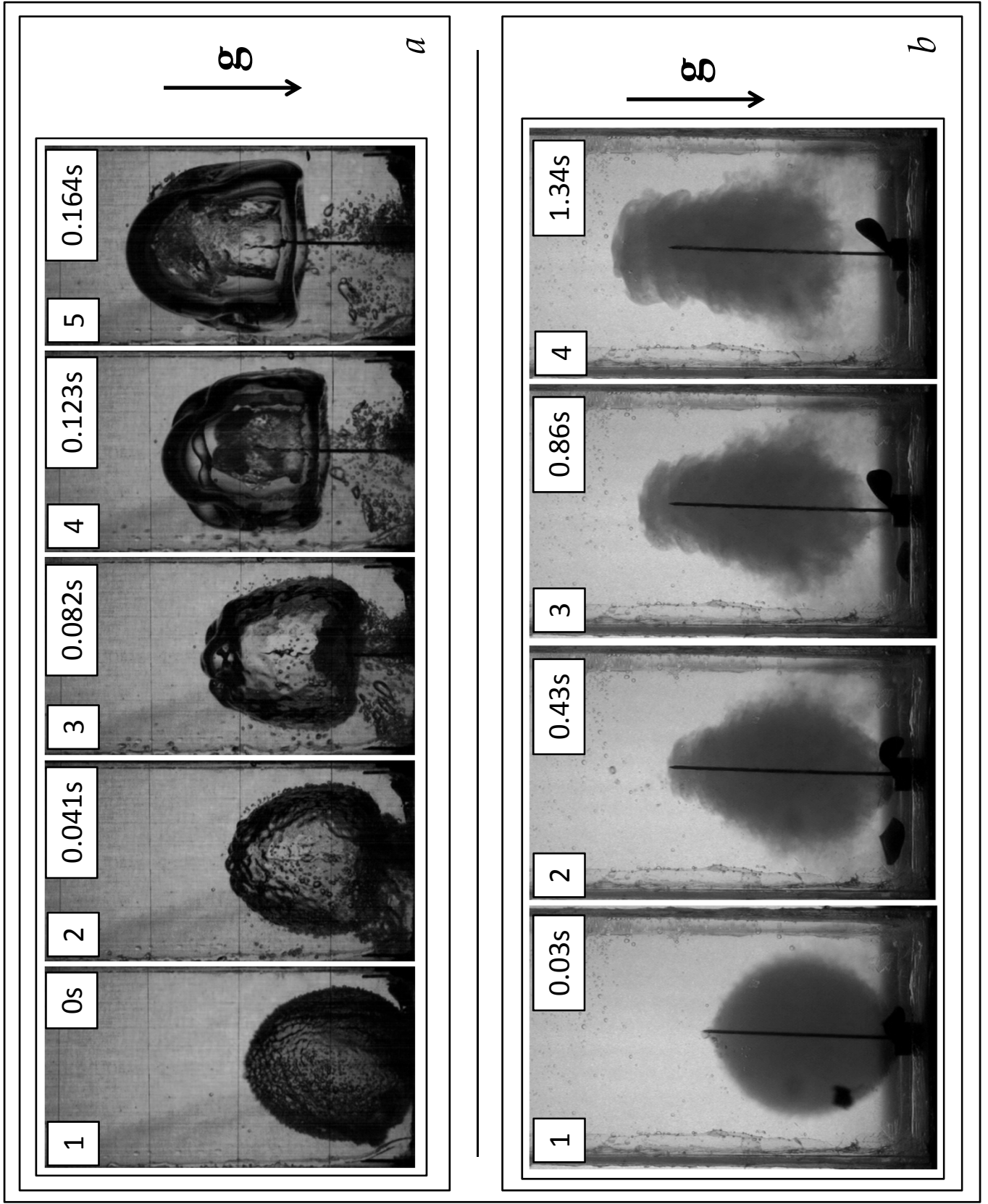